\documentstyle[pra,aps,psfig]{revtex}
\newcommand{\beq}{\begin{equation}}
\newcommand{\eeq}{\end{equation}}
\newcommand{\beqa}{\begin{eqnarray}}
\newcommand{\eeqa}{\end{eqnarray}}
\newcommand{\ba}{\begin{array}}
\newcommand{\ea}{\end{array}}

\begin{document}
\draft

\twocolumn[\hsize\textwidth\columnwidth\hsize\csname
@twocolumnfalse\endcsname

\widetext 
\title{Colored Noise in Quantum Chaos} 
\author{Luca Salasnich} 
\address{Istituto Nazionale per la Fisica della Materia, 
Unit\`a di Milano Universit\`a \\ 
Dipartimento di Fisica, Universit\`a di Milano, \\
Via Celoria 16, 20133 Milano, Italy} 

\maketitle

\begin{abstract} 
We derive a set of spectral statistics whose power spectrum 
is characterized, in the case of chaotic quantum systems,  
by colored noise $1/f^{\gamma}$, where the integer parameter 
$\gamma$ critically 
depends on the specific energy-level statistic considered. 
In the case of regular quantum systems these spectral statistics 
show $1/f^{\gamma+1}$ noise. 
\end{abstract} 


]

\narrowtext 

Quantum chaos is the study of quantum systems which are 
classically chaotic \cite{chaos-book}. There is a conjectured relationship, 
backed up by numerous examples, between the energy level fluctuation 
properties of a quantum systems and the dynamical behavior of 
its classical analog. In particular, Berry and Tabor showed 
that classically integrable systems are characterized by 
Poissonian spectral fluctuations \cite{berry}, while Bohigas, 
Giannoni and Schmit \cite{bohigas} suggested that 
classically chaotic systems can be described by 
Gaussian ensembles of random matrix theory (RMT). In the past years 
many energy-level statistics have been proposed 
to characterize the spectral fluctuations; among them there are 
the nearest-neighbour level-spacing distribution $P(s)$, 
the spectral rigidity $\Delta_3(L)$ and 
the spectral form factor $K(\tau )$. 
Recently, Relano {\it et al.} \cite{sp1} have introduced 
the energy-level statistic $\delta_n$, 
showing that chaotic quantum systems display $1/f$ noise 
in the $\delta_n$ statistic, whereas integrable ones exibit 
$1/f^2$ noise \cite{sp1,sp2}. 
\par 
In this Brief Report we introduce a new set of spectral statistics 
which contains the spectral statistic $\delta_n$ 
as a particular case. We show that 
for chaotic quantum systems these statistics are 
characterized by colored noise $1/f^{\gamma}$, 
where the integer parameter $\gamma$ critically 
depends on the specific energy-level statistic considered. 
Moreover, we find that in the case of regular quantum systems 
these spectral statistics display $1/f^{\gamma+1}$ noise. 
\par 
Let us consider a quantum system with a discrete set of unfolded levels 
$E_n$, such that the energy is measured in units of the mean level speacing.  
The staircase function $N(E)$, which gives the number of energy levels 
up to the energy $E$, can be written as 
\beq 
N(E) = {\bar N}(E) + N_{osc}(E) \; , 
\eeq 
where ${\bar N}(E)=E$ is the averaged number of levels 
and $N_{osc}(E)$ is the fluctuating part. $N_{osc}(E)$ is supposed 
to belong to a universality class, which should only depend on 
the integrability or chaoticity of the classical analog \cite{chaos-book}. 
The Madrid group of Relano {\it et al.} \cite{sp1} has analyzed these 
fluctuations by using the spectral statistic $\delta_n$, defined as 
\beq 
\delta_n = N_{osc}(E_{n}) \; . 
\eeq 
In particular, the Madrid group has investigated 
the behavior of the power spectrum $S(k)$ of the discrete 
series $\delta_n$, given by  
\beq 
S_M(k) = |{\hat \delta}_k |^2 \; , 
\eeq 
where ${\hat \delta}_k$ is the Fourier transform of $\delta_n$, 
\beq 
{\hat \delta}_k = {1\over \sqrt{M}} \sum_{n=1}^M \delta_n 
\exp\left({- i k n \over M} \right) \; , 
\eeq 
and $M$ is the size of the series. By studying spectra of atomic 
nuclei at high energies and also spectra of Gaussian ensebles 
of RMT, the Madrid group has found the power law 
$\langle S_M(k) \rangle \sim 1/k$. 
For Poisson spectra it has found instead 
$\langle S_M(k)\rangle \sim 1/k^2$, 
as expected for independent random variables. 
Finally, the following conjecture has been suggested: the energy spectra 
of chaotic quantum systems are characterized by pink noise $1/f$, 
in contrast to the Brown noise $1/f^2$ of regular systems \cite{sp1}. 
\par 
The Madrid group has analytically proved its conjecture  
on the basis of the RMT \cite{sp2}. Independently, Robnik has 
obtained the same result \cite{robnik}: 
under the conditions $M\gg 1$ and $k/M\ll 1$, 
the averaged power spectrum of the energy-level statistic 
$\delta_n$ is given by 
\beq 
\langle S_M(k) \rangle 
= \left\{  \ba{cc} 
{2 \over \beta} {M\over k} & \mbox{for chaotic systems} \\ \\
{M^2\over k^2} & \mbox{for integrable systems} \\
\ea 
\right. \; , 
\eeq 
where $\beta$ depends on the symmetry of the Gaussian ensemble: 
$\beta=1$ for the Gaussian Ortogonal Ensemble (GOE), $\beta=2$ for the 
Gaussian Unitary Ensemble (GUE), and $\beta=4$ for the Gaussian Symplectic 
Ensemble (GSE). Thus, although the origin of $1/f$ noise 
in many complex systems 
is still an unsolved problem \cite{mandelbrot}, 
the origin of $1/f$ noise in the spectral 
fluctuations of chaotic quantum systems has been easier to 
understand because of the
mathematical tractability of RMT. 
It is important to stress that Eq. (5) is valid if the condition 
$k/M\ll 1$ is fullfilled. In fact, as shown in \cite{sp2} 
the Eq. (5) is the power-law approximation, obtained 
via Taylor expansion, of a more complex formula. 
In Fig. 1 we plot both the exact curve derived in \cite{sp2} 
and the power-law curve of Eq. (5). Figure 1 
shows that $\langle S_M(k) \rangle$ presents 
deviations from the power-law behavior 
for $k$ close to the Nyquist number $k_c=M/2$. These deviations are due 
to the finite size $M$ of the series $\delta_n$ and the figure 
shows that they are negligible for $k/M\le 10^{-2}$, where 
the relative error is below $1\%$. In practice, 
because of the log-log scale, the differences are 
clearly visible only for $k/M>10^{-1}$.  

\begin{figure}
\centerline{\psfig{file=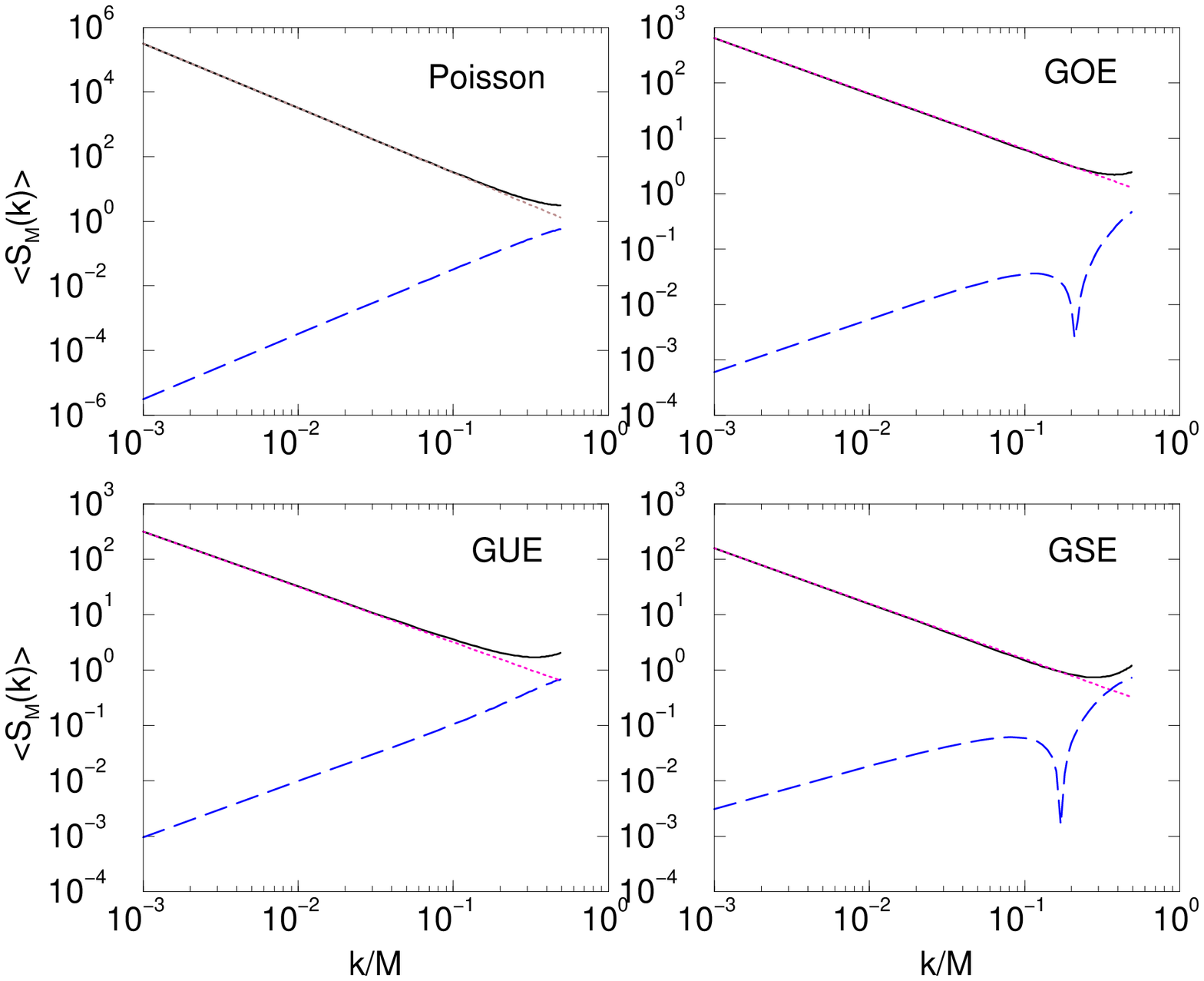,height=2.8in}}
\small 
{FIG. 1 (color online). Average power spectrum $\langle S_M(k) \rangle$ 
of the spectral statistic $\delta_n$. Comparison between the 
exact formula of Ref. [5] (solid line) and its 
power-law approximation (dotted line), given by Eq. (5). 
The dashed line is the relative error of the two curves. 
Note the different scale of the Poisson ensemble 
(integrable systems) with respect to the 
Gaussian ensembles (chaotic systems).} 
\end{figure} 

\par 
The proof of the nice result of Eq. (5) is based on the relationship 
between $S_M(k)$ and the widely studied spectral form factor 
of the density of levels \cite{sp2,robnik}. Here, 
we apply the same methodology of Ref. \cite{sp2,robnik} 
to introduce and analyze a new set of spectral statistics, 
which contains $\delta_n$ 
as a particular case. First we define the following set of functions 
\beq 
N_{osc}^{(\alpha )}(E) = {d^{\alpha} N_{osc}(E)\over dE^{\alpha}  } \; ,  
\eeq 
where the parameter $\alpha$ is a positive integer. This family of functions 
can be extended also to negative integer values of $\alpha$ by setting 
\beq 
N_{osc}^{(-|\alpha |)}(E) = 
\int_{-\infty}^E dx_1 \int_{-\infty}^{x_1} dx_2 ... 
\int_{-\infty}^{x_{|\alpha| -1}} dx_{|\alpha|} N_{osc}(x_{|\alpha|} )  \; . 
\eeq 
For $\alpha = 0$ one recovers the oscillating part $N_{osc}(E)$ 
of the staircase function, while for $\alpha = 1$ one has the 
oscillating part $\rho_{osc}(E)$ of the density of levels. In analogy 
with Eq. (2) we now introduce the following set of 
spectral statistics:  
\beq 
\delta_n^{(\alpha )} = N_{osc}^{(\alpha )}(E_{n}) \; ,  
\eeq 
spanned by the integer number $\alpha$. 
Obviously, for $\alpha = 0$ one has the 
spectral statistic $\delta_n$ of Eq. (2). 
\par 
By applying the formula of integration by parts, 
it is straightforward to prove that 
the Fourier transform of $N_{osc}(E)$, given by 
\beq 
{\hat N}_{osc}(t) = 
\int_{-\infty}^{+\infty} 
N_{osc}(E) \exp\left( - i E t \right) \; dE \; , 
\eeq
is related to the Fourier transform of $N_{osc}^{(\alpha )}(E)$ 
by the simple formula 
\beq 
{\hat N}_{osc}(t) = {\hat N}_{osc}^{(\alpha)}(t) {1\over (i t)^{\alpha} } \; . 
\eeq 
Under the conditions $M\gg 1$ and $k/M\ll 1$ the discrete 
Fourier transform of $\delta_n$ can be obtained from the 
Fourier transform of $N_{osc}(E)$. It is given by  
\beq 
{\hat \delta}_k = {\hat N}_{osc}({k \over M}) \; . 
\eeq
Taking into account Eqs. (3), (8), (10) and (11), 
the relationship between the power spectrum of 
$\delta_n$ and the power spectrum of $\delta_n^{(\alpha)}$ reads 
\beq 
S_M(k) = { S_M^{(\alpha)}(k) M^{2\alpha} \over k^{2\alpha} } \; . 
\eeq 
Finally, by using the Eq. (5) one finds that 
the average power spectrum of the spectral statistic $\delta_n^{(\alpha)}$ 
satisfies the following formula 
\beq
\langle S_M^{(\alpha)}(k) \rangle = \left\{  
\ba{cc}
{2 \over \beta} {M^{1-2\alpha} 
\over k^{1-2\alpha} } & \mbox{for chaotic systems} \\ \\
{M^{2-2\alpha} \over k^{2-2\alpha} } & \mbox{for integrable systems} \\
\ea 
\right. 
\eeq 
This formula is the main result of the paper. For $\alpha =0$ it reduces 
to Eq. (5). In general, for $\alpha >0$ the average 
power spectrum of $\delta_n^{(\alpha)}$ 
is not divergent, it is instead divergent for $\alpha \leq 0$. 
\par 
Among the new spectral statistics, for numerical purposes 
the most simple ones are $\delta_n^{(-1)}$ and $\delta_n^{(1)}$. 
The statistic $\delta_n^{(-1)}$ is given by  
\beq 
\delta_n^{(-1 )} =  \int_{-\infty}^{E_n} N_{osc}(E) \; dE \; ,  
\eeq 
and its average power spectrum is 
\beq
\langle S_M^{(-1)}(k) \rangle = \left\{  
\ba{cc}
{2 \over \beta} {M^{3} \over k^{3} } & \mbox{for chaotic systems} \\ \\
{M^{4} \over k^{4} } & \mbox{for integrable systems} \\
\ea 
\right. 
\eeq 
Thus, $\delta_n^{(-1)}$ shows black noise $1/f^3$ for chaotic systems 
and black noise $1/f^4$ for regular systems.  
The statistic $\delta_n^{(1)}$ is instead given by 
\beq 
\delta_n^{(1)} = {dN_{osc}(E_{n})\over dE} = \rho_{osc}(E_{n}) 
\; .   
\eeq 
Its average power spectrum is nothing else than 
the discrete spectral form factor of the density of levels, 
which shows blue noise $f$ for chaotic systems and 
white noise $f^0=1$ for regular systems. 

\begin{figure}
\centerline{\psfig{file=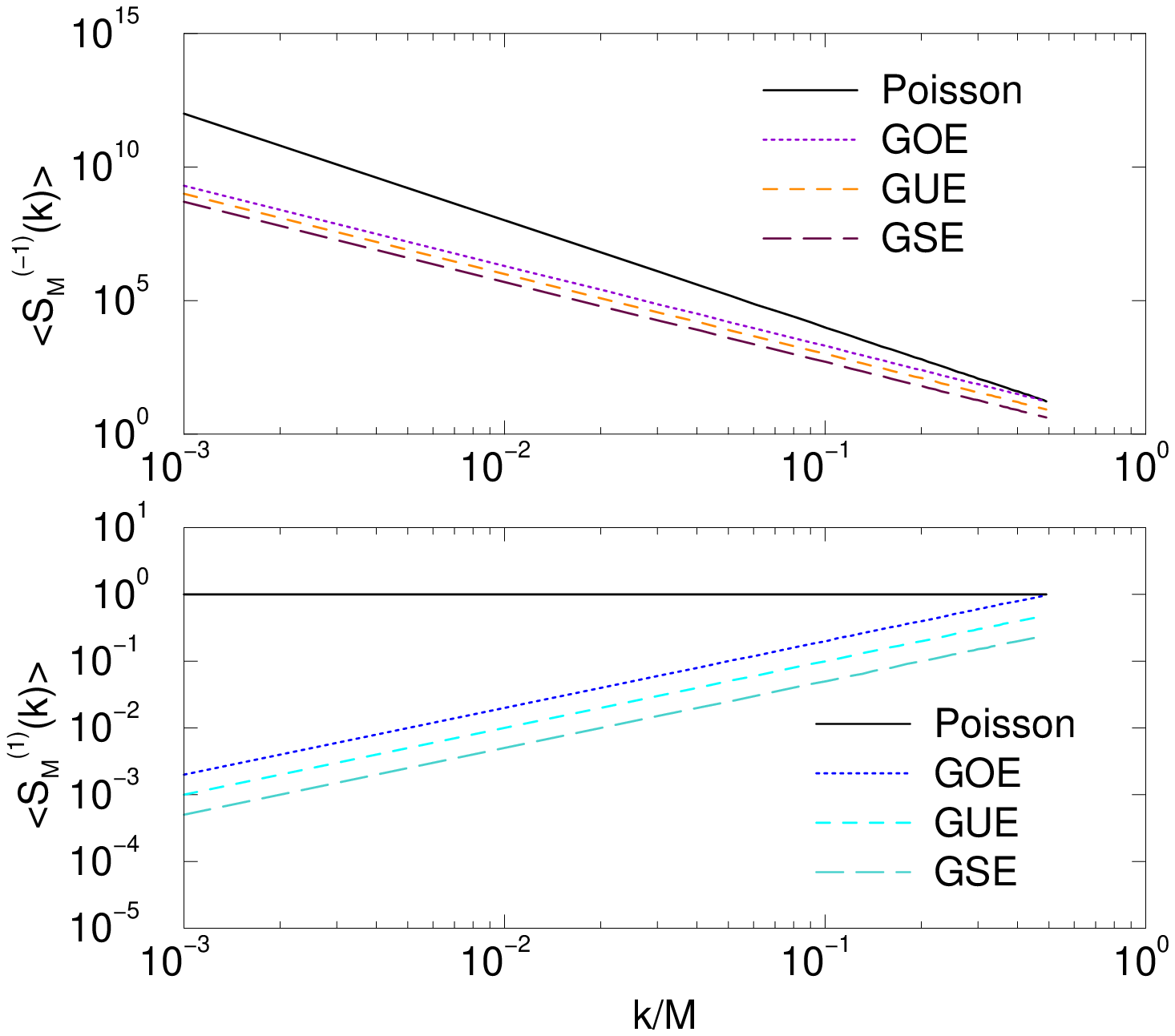,height=2.8in}}
\small 
{FIG. 2 (color online). 
Average power spectrum $\langle S_M^{(\alpha)}(k) \rangle$ 
of the spectral statistics $\delta_n^{(-1)}$ 
and $\delta_n^{(1)}$. Power laws for integrable 
systems (Poisson) and for chaotic systems with 
different classes of symmetry: GOE, GUE, and GSE.} 
\end{figure} 

For the sake of completeness, in Fig. 2 we plot the power 
spectrum of spectral statistics $\delta_n^{(-1)}$ (upper panel) 
and $\delta_n^{(1)}$ (lower panel). Figure 2 shows that 
different power laws can be easily distinguished 
by choosing the appropriate scale interval in the log-log plot. 
The new spectral statistics we have introduced are a simple and 
useful tool to analyze the energy levels of quantum systems and 
determine their integrability or chaoticity. 
\par 
An open problem is the behavior of $\langle S_M^{(\alpha )}(k)\rangle$ 
for systems in the mixed regime between order and chaos. 
Recently we have analyzed the order to chaos transition 
in terms of the power spectrum 
$\langle S_M(k)\rangle$ by using the Pascal's snail 
(Robnik billiard) \cite{billiard}. We have numerically found a net 
power law $1/f^{\gamma}$, with $1\le \gamma \le 2$, 
at all the transition stages. We have suggested that 
the exponent $\gamma$ is related to the chaotic component 
of the classical phase space of the quantum billiard, but 
a theoretical explanation of these numerical results 
is still lacking \cite{joint}. 
\par 
In conclusion, we have shown that colored noise $1/f^{\gamma}$ 
characterizes the spectral fluctuations of quantum systems. 
The presence of colored noise in fluctuating physical variables 
is ubiquitous. This kind of noise has been detected 
in DNA sequences, quasar emission, river discharge, heartbeat, 
among many others. Despite this ubiquity, there is no universal 
explanation about this phenomenon, which is a generic manifestation 
of complex systems. Nevertheless, here we have shown that 
for integrable and chaotic quantum systems 
the colored noise of energy-level fluctuations 
can be explained in a satisfactory way 
on the basis of random matrix theory. 
\par 
The author thanks J.M.G. Gomez, A. Relano, E. Faleiro 
and M. Robnik for many useful discussions.

\end{document}